\documentclass[useAMS,usenatbib]{mnras}

\usepackage{graphicx}
\usepackage{amssymb}
\usepackage{amsmath}
\newcommand{\PWVW}{PWV$_{\rm{W}}$}
\newcommand{\PWVII}{PWV$_{\rm{W24}}$}
\newcommand{\PWVIV}{PWV$_{\rm{W48}}$}
\newcommand{\PWVG}{PWV$_{\rm{G}}$}

\title[PWV forecasting validation]{Precipitable water vapour forecasting: a tool for optimizing IR observations at Roque de los Muchachos Observatory.}

\author[G. P\'erez-Jord\'an et al.]
  {G.~P\'erez-Jord\'an,$^1$\thanks{Visiting fellow.}\thanks{E-mail: gpj@iac.es (GPJ); jcastro@iac.es (JACA); cmt@iac.es (CMT)}
  J.A~Castro-Almaz\'an,$^{1,2}$\footnotemark[2] C.~Mu\~noz-Tu\~n\'on,$^{1,2}\footnotemark[2]$ 
  \\
	$^1$Instituto de Astrof\'isica de Canarias, E-38200, La Laguna, Spain\\
	$^2$Dept. Astrof\'isica, Universidad de La Laguna, E-38200, La Laguna, Spain}

\begin{document}
\maketitle

\label{firstpage}

\begin{abstract}
\label{sec:abstract}
We validate the Weather Research and Forecasting (WRF) model for precipitable water vapour (PWV) forecasting as a fully operational tool for optimizing astronomical infrared (IR) observations at Roque de los Muchachos Observatory (ORM). For the model validation we used GNSS-based (Global Navigation Satellite System) data from the PWV monitor located at the ORM. We have run WRF every $24$ h for near two months, with a horizon of $48$ hours (hourly forecasts), from 2016 January 11 to 2016 March 4. These runs represent $1296$ hourly forecast points. The validation is carried out using different approaches: performance as a function of the forecast range, time horizon accuracy, performance as a function of the PWV value, and performance of the operational WRF time series with 24- and 48-hour horizons. Excellent agreement was found between the model forecasts and observations, with R $=0.951$ and R $=0.904$ for the 24- and 48-h forecast time series 
respectively. The $48$-h forecast was further improved by correcting a time lag of $2$ h found in the predictions. The final errors, taking into account all the uncertainties involved, are $1.75$ mm for the $24$-h forecasts and $1.99$ mm for $48$ h. We found linear trends in both the correlation and RMSE of the residuals (measurements $-$ forecasts) as a function of the forecast range within the horizons analysed (up to $48$ h). In summary, the WRF performance is excellent and accurate, thus allowing it to be implemented as an operational tool at the ORM.

\end{abstract}


\section{Introduction and objectives}
\label{sec:intro}
In a previous paper \citep{per15} we validated the Weather Research and Forecasting (WRF) Numerical Weather Prediction (NWP) model for the precipitable water vapour (PWV) at astronomical sites. We used high resolution radiosonde balloon data launched at Roque de los Muchachos Observatory (ORM) in the Canary Islands and, from a comparison, we proposed a calibration for the highest horizontal resolution ($3$ km) results. Abundant literature exists addressing the success of mesoscale NWP models in PWV forecasting \citep{cuc00, mem05, zhu08, cha10, poz11, gon13, poz16}. Some of these studies are centred on the use of WRF at the ORM \citep{per10, gon13, per15}. \citet{gio13} also tested the model for meteorological and optical turbulence conditions and, in a subsequent paper, \citet{gio14} applied WRF at the ORM to validate the model as a possible tool in examining potential astronomical sites all over the world.

Although water vapour (WV) represents only about $3.3 \times 10^{-3}$ per cent of the atmosphere's total mass, it is the main absorber at IR, millimetre, and submillimetre wavelengths; it is also an important source of the thermal IR background. WV can be assessed through the PWV value, defined as the total amount of WV contained in a vertical column of unit cross-sectional area from the surface to the top of the atmosphere. PWV is commonly expressed in mm, meaning the height that the water would reach if condensed and collected in a vessel of the same unit cross-section.
Generally speaking, the vertical distribution  of PWV decreases with height but shows high spatial and temporal variability \citep{ota11}. It  is also important to emphasize that for the ORM, the PWV content cannot be described merely as a function of altitude \citep{ham98}; other factors, such as the thickness of the troposphere, have also to be considered \citep{gar04}. 

\begin{figure}
	\includegraphics[width=\columnwidth]{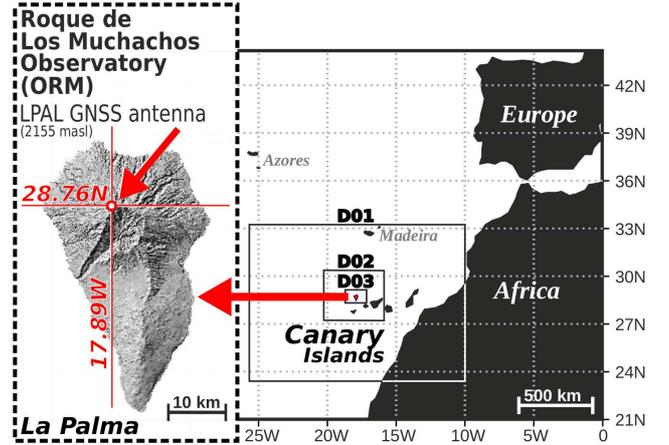} 
	\caption{Location of Roque de los Muchachos Observatory (ORM) and the LPAL geodetic GNSS antenna on the island of La Palma (Canary Islands), together with the nested domains used for the WRF forecasts (see Table \ref{table:domains} for details).}
	\label{fig:map}
\end{figure}

The ORM, in La Palma (Canary Islands, Spain), is listed among the first-class astronomical sites worldwide. The latitude of the islands and their location in the eastern North Atlantic Ocean (see Fig.\ \ref{fig:map}), together with the cold oceanic stream, define the characteristic vertical troposphere structure with a trade wind thermal inversion layer (IL), driven by subsiding cool air from the descending branch of the Hadley cell. The altitude of the IL ranges on average from $800$ m in summer to $1600$ m in winter, well below the altitude of the ORM \citep{dor96,car16}. The IL separates the moist marine boundary layer from the dry free atmosphere, inducing high atmospheric stability above it and low values of PWV \citep{gar10}. The Observatory covers an area of $189$ hectares and hosts an extensive fleet of telescopes, including the largest optical-IR telescope to date, the $10.4$ m Gran Telescopio Canarias (GTC). The GTC has three IR instruments\footnote{http://www.gtc.iac.es/instruments}: CIRCE 
and EMIR (in the $JHK$ bands, 1--2.5$~\mu$m) and CanariCam operates at longer wavelengths (10--20~$\mu$m).

The PWV content determines whether or not IR observations are feasible. Observations at longer wavelengths (such as those with CanariCAM) are even more restrictive in their PWV requirements. PWV below $3$ mm is a reference value for observations to be scheduled for this instrument. In this sense the ORM, which manages to sustain these conditions for $\approx40\%$ of the time \citep{gar10} has proven to be most suitable. However, the prevailing PWV value is not the only parameter that defines the suitability of a site for IR observations. Knowledge of the local trend and temporal stability are also critical in determining the efficiency of observing in the IR, in terms of both the availability of time and the practicality of scheduling the telescope to exploit this time.

{\it A priori} knowledge of this atmosphere parameter enables us to get the most from an observing site. In particular, the possibility of knowing the PWV value in advance is mandatory in scheduling queue mode operation in IR astronomy. The aim of the present paper is to validate WRF as a fully operational tool for optimizing astronomical IR observations at the ORM by characterizing its performance, and quantifying the its accuracy and operational capabilities. To achieve this objective, we have included, for comparison, data from a PWV time series measured at the ORM (see Section \ref{sec:GNSS_ts}) with a monitor based on the Global Navigation Satellite System (GNSS; Global Positioning System, GPS) technique \citep{bev92,bev94} with input data from a permanent antenna (LPAL, see Fig. \ref{fig:map}).

This paper is structured as follows. Sections \ref{sec:WRF_model} and \ref{sec:GNSS_ts} describe the WRF model and the PWV GNSS monitor. Section \ref{sec:PWV_datasets} presents the datasets. The results of the comparison between the PWV values forecast by WRF and measured with the GNSS monitor are given in Section \ref{sec:results}. In Section \ref{sec:PWV_gradients} there is a brief discussion of the ability of WRF to forecast steep PWV variations. Finally, Section \ref{sec:operational_tool} discusses the practical aspects of WRF as an operational tool for PWV forecasting in an astronomical 
context.

\section{The WRF model}
\label{sec:WRF_model}

WRF is a non-hydrostatic mesoscale meteorological model designed for research and operational applications \citep{ska08}. It was developed as a collaboration between various US institutions: the National Center for Atmospheric Research (NCAR), the National Oceanic and Atmospheric Administration (NOAA), the Air Force Weather Agency (AFWA), the Naval Research Laboratory (NRL), the University of Oklahoma (OU), and the Federal Aviation Administration (FAA). In contrast with global models, a mesoscale meteorological model has higher horizontal and vertical resolution so that it can better represent the subgrid processes, especially in areas with abrupt orography, such as the Canary Islands. Moreover, WRF offers improved time resolution in the forecast variables, and an ample set of configuration options is available. The model \emph{domain} covers a vast mesoscale area that has to be solved with an appropriate selection of 
initial conditions for the input variables, including temperature ($T$), relative humidity (RH), and the $U$ and $V$ components of wind velocity, in $\approx32$ vertical levels.\footnote{The vertical levels in the external GFS files are: surface, 1000, 975, 950, 925, 900, 850, 800, 750, 700, 650, 600, 550, 500, 450, 400, 350, 300, 250, 200, 150, 100, 70, 50, 30, 20, 10, 7, 5, 3, 2, and 1 hPa.} In this study, we obtain the initial conditions from the Global Forecast System\footnote{http://www.emc.ncep.noaa.gov} (GFS), a global model produced by the US National Centers for Environmental Prediction (NCEP). Once the first domain is solved, a recursive horizontal grid-nesting process focuses on the area of interest with the required horizontal resolution. The physical domain in WRF is set with the WRF Preprocessing System (WPS) module. The domain configuration (see Fig. \ref{fig:map} and Table \ref{table:domains}) 
is summarized as:

\begin{itemize}
    \item[$-$] A coarse domain with horizontal resolution $\Delta x=27$ km (D01).
    \item[$-$] Two consecutive nests with  horizontal resolutions $\Delta x=9$ km (D02) and $\Delta x=3$ km (D03).
	\item[$-$] A grid-distance ratio of 3:1 for domain nesting.
    \item[$-$] Thirty-two vertical levels, with separations ranging from $\sim$100 m, close to the surface, to $\sim$1500 m, near the tropopause ($\approx$14 km).
\end{itemize}

\begin{table}
 \centering
  \caption{Configuration of the nested domains used for the WRF forecasts (see Fig. \ref{fig:map}).}
 \label{symbols}
 \begin{tabular}{@{}ll cc r}
   \hline
   Domain & $\Delta x$ (km) & Grid & Surface (km) & Surface (degrees)\\
   \hline
   D01 & 27 & $60\times45$ & $1620\times1215$ & $14.7\degr \times 11\degr$ \\
   D02 & 9  & $52\times40$ & $468\times360$ & $4.2\degr \times 3.3\degr$ \\
   D03 & 3  & $40\times25$ & $120\times75$ & $11.1\degr \times 0.7\degr$ \\
   \hline
   \label{table:domains}
 \end{tabular}
\end{table}

The WRF equations are formulated using a vertical coordinate defined as:
\begin{equation}
 \label{eq:eta}
 \eta=\frac{p_{z}-p_{\rm top}}{p_{\rm s}-p_{\rm top}},
\end{equation}
where {$p_{\rm top}$} is the pressure at the model's top level, {$p_{\rm s}$} is the surface pressure, and {$p_{z}$} is the pressure at any level $z$. All the values refer to the hydrostatic component of pressure. The surface inputs make $\eta$ a \emph{terrain-following} variable. The value of $\eta$ ranges from 1 at the surface to 0 at the upper boundary of the vertical domain, which we have fixed at $10$ hPa. The vertical level configuration may be customized by the user.

The subgrid scale processes occur at scales too small to be explicitly resolved by the model, so they are parametrized through the \emph{physics} of the model. Model physics in WRF is implemented in different modules: Microphysics, Radiation (Short-Wave -- SW -- and Long-Wave -- LW), Cumulus, Surface Layer (SL), Land-Surface (LS), and Planetary Boundary Layer (PBL). WRF permits the selection of different schemes for each physics module. In particular, the LS schemes provide heat and moisture fluxes acting as a lower boundary condition for the vertical transport carried out in the PBL schemes. The PBL scheme assumes that the BL eddies cannot be resolved with analytical equations and includes a set of empirical parametrizations. This is a key point, as the BL eddies are responsible for vertical subgrid scale fluxes due to energy transport in 
the whole atmospheric column, not just in the BL. In WRF, the PBL schemes are divided into two categories: non-local and Turbulent Kinetic Energy (TKE) local schemes.

\subsection{Initial and boundary conditions}
\label{sec:WRF_IBC}

As mentioned previously, in order to start the integration of the dynamical equations in WRF, initial and boundary conditions (IBC) are needed. The IBC can be obtained from an external analysis or forecast interpolated to the WRF grid points. The WPS module processes the IBC to generate the meteorological and terrestrial data inputs for WRF. In this work we use GFS to feed WRF. We carried out different experiments to show that the best correlation with the observed data is that with the highest available GFS frequency and resolution, i.e.\ every 3 hours and $0.25\degr\times0.25\degr$ (upgraded in January 2015).

\subsection{Configuration}
\label{sec:WRF_conf}

\begin{table*}
 \centering
  \caption{WRF model physics configuration. The selected scheme is shown under each model physics module.}
 \label{symbols}
 \begin{tabular}{@{}llll llll}
   \hline
   LW Radiation & SW Radiation & Radiation timestep & Land Surface & Surface Layer & PBL & Cumulus & Microphysics\\
   \hline
   RRTM & Dudhia & 27 & Noah LSM & Monin-Obukhov & YSU & Kain-Fritsch & WSM6\\
   \hline
   \label{table:physics}
 \end{tabular}
\end{table*}

\begin{figure}
	\includegraphics[width=84mm]{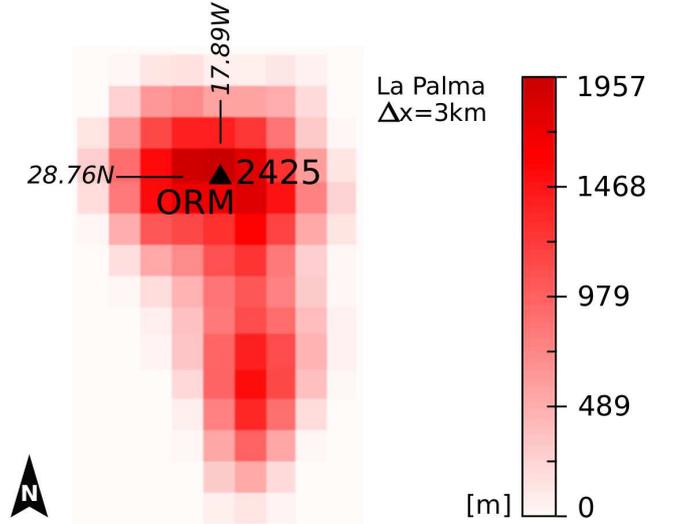} 
	\caption{La Palma as seen by WRF with a horizontal resolution of $\Delta x=3$ km (D03 domain; see Fig.\ \ref{fig:map}). Owing to the steep orography, the pixel that includes the ORM (maximum altitude = $2425$ m) extends northwards to the downward slope, with an average altitude of $1957$ m. This altitude is lower than that of the LPAL GNSS antenna (see Fig.\ \ref{fig:sketch}).}
	\label{fig:lapalma_3km}
\end{figure}

\begin{figure}
	\includegraphics[width=84mm]{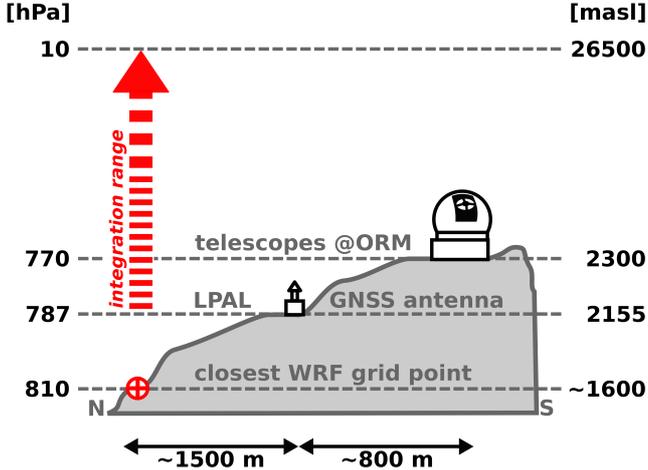} 
	\caption{Schematic sketch showing the local distances between the telescope locations at the ORM, the LPAL GNSS antenna used by the PWV monitor, the closest WRF grid point in the D03 domain, and the vertical range used for the integration of PWV (from $787$ hPa to $10$ hPa).}
	\label{fig:sketch}
\end{figure}

WRF supports different projections on the sphere. We have selected the Mercator projection as it is best suited for low latitudes and also because of the predominant west--east extent of our domains. Under this projection, the \emph{true latitude}, at which the surface of projection intersects (or is tangential to) the surface of the Earth (no distortion point), has been set to $30\degr$N. The three nested domains D01, D02, and D03 (see Fig.\ \ref{fig:map} and Table \ref{table:domains}) have all been configured to be centred on a coordinate point at the ORM ($28\degr 45.5\arcmin$ N, $17\degr 52.5\arcmin$ W). USGS (US Geological Survey) geographical data was used to set up the model domains with resolutions of $10\arcmin$, $2\arcmin$, and $0.5\arcmin$ for D01, D02, and D03, respectively. This means that the precision of the geographical coordinates is limited to $\approx$900 m for the best case in the D03 domain. In Fig.\ \ref{fig:lapalma_3km} we have shown the effect of the horizontal resolution ($\Delta 
x=3$ km) on the geographical altitude model. Owing to the steep orography, the pixel that includes the ORM extends northwards to the downward slope, with an average altitude of $1957$ m. This altitude is lower than the level at which the LPAL GNSS antenna is located ($2155$ m; see Fig.\ \ref{fig:sketch}). This effect is corrected by the trimming lower limit of the integration range to obtain PWV to the closest mean pressure level of the antenna ($\approx$787 hPa) and by applying a local calibration to the data (see Sec.\ \ref{sec:PWV_datasets}). Once the three nested domains were centred on the ORM, we selected the closest D03 WRF grid point to run the model ($28\degr 46.5\arcmin$ N, $17\degr 52.5\arcmin$ W). This point is $\approx$1.5 km NE of the GNSS antenna, at an altitude of $\approx$1600 m (see Fig.\ \ref{fig:sketch}). Regarding the way the nested domains interact, WRF supports various options. We have selected the two-way nesting, in which the fine domain (D03) solution replaces the coarse domain (D02)
 solution for the grid points of D02 that lie inside D03.

The model physics configuration is listed in Table \ref{table:physics} and is summarized as follows:

\begin{itemize}
    \item[$-$] The Rapid Radiative Transfer Model (RRTM) and Dudhia have been selected for LW and SW radiation respectively.
    \item[$-$] For cumulus parametrization we used the Kain--Fristch (new Eta) scheme, which uses a relatively complex cloud model for horizontal resolutions $\geq9$ km. Below this value, we assume that the convection is reasonably well resolved by the non-hydrostatic component of the WRF dynamics.
    \item[$-$] The Noah--LSM scheme has been selected as the Land Surface scheme. It is well tested and includes snow cover prediction.
    \item[$-$] A widely used nonlocal scheme (Yonsei University or YSU) has been selected for the PBL.
    \item[$-$] The Monin--Obukhov scheme Surface Layer is used (SL). In version 3 of WRF each PBL scheme must use a specific SL scheme.
    
\end{itemize}

\section{PWV datasets and methods}
\label{sec:PWV_datasets}

In this study we are using two PWV datasets, one forecast by the WRF model and one measured with a GNSS monitor. Both time series cover a period of near two months, from 2016 January 11 to 2016 March 4. 

\subsection{The WRF time series}
\label{sec:WRF_ts}

We ran WRF (see Sec.\ \ref{sec:WRF_model}) every $24$ h (at $12$ UTC) with a horizon of $48$ hours for the two months studied. Therefore, the full dataset, \PWVW~hereafter, includes a total of $54$ WRF simulations with $1296$ hourly points forecast twice: one with a prediction horizon up to $48$ h (\PWVIV) and the other (the next day run) with a prediction horizon up to $24$ h (\PWVII). 

The data were calibrated using the equation obtained by \cite{per15} directly at ORM for the resolution $\Delta x = 3\rm{km}$ (D03 domain), after a validation with local high resolution radiosonde balloons with correlation $=0.970$:

\begin{equation}
\label{eq:calib_d03}
PWV_{\rm{W}}=1.01\cdot PWV_{D03} - 0.82;~(\rm{RMSE = 0.82~mm}),
\end{equation}

\noindent where $PWV_{D03}$ is the raw output of WRF for the domain D03. 

\subsection{The GNSS time series}
\label{sec:GNSS_ts}

As a valid reference for comparison and validation, we included a simultaneous series of PWV from the GNSS monitor at the ORM. The technique for retrieving PWV from the tropospheric delays induced in the GNSS signals has been explained, for example, by \citet{bev92,bev94,gar10,cas16}. The delays result from the difference in the refracted and straight line optical paths, that can be derived after a least-squares fit of the signals received from a constellation of $\approx$10 satellites over a typical two-hour average lag \citep{bev92}. The total delay, projected to the zenith and corrected for the ionospheric component (tropospheric zenith delay, TZD), may be separated into two terms, the zenith hydrostatic delay (ZHD), which changes slowly and can be modelled as a function of the local barometric pressure ($p_s$), the latitude ($\phi$) and the altitude ($h$) of the antenna \citep{elg91}, and the zenith wet delay (ZWD) \citep{saa72}, which is directly proportional to the PWV \citep{ask87}.

Spain's Instituto Geogr\'afico Nacional (IGN) maintains the geodetic GNSS antenna LPAL next to the ORM residential buildings as part of the EUREF Permanent GNSS Network\footnote{http://www.epncb.oma.be} (see Figs\  \ref{fig:map} and \ref{fig:sketch}). The IAC has developed\footnote{subcontractor: Soluciones Avanzadas Canarias} an online PWV monitor based on the GNSS data from LPAL\footnote{www.iac.es/site-testing/PWV\_ORM} \citep{gar10,cas16} with a temporal resolution of $0.5$ h. This frequency allows us to test the temporal accuracy of WRF in forecasting episodes with abrupt changes in PWV. The series (hereafter \PWVG ) were subsampled to a frequency of $1$ h to match with \PWVW, and were calibrated using the equation obtained by \cite{cas16} for this monitor after a validation with operational radiosonde balloons launched from the neighbouring island of Tenerife with a correlation $=0.970$:

\begin{equation}
\label{eq:calib_gnss}
PWV_{\rm{G}}=0.97\cdot PWV_{G_{raw}} - 1.39;~(\rm{RMSE = 0.70~mm}).
\end{equation}

There is a gap in \PWVG~from February 18 to February 22 because of a PWV monitor outage caused by an intense snow storm that covered the antenna.

\begin{figure}
	\includegraphics[width=84mm]{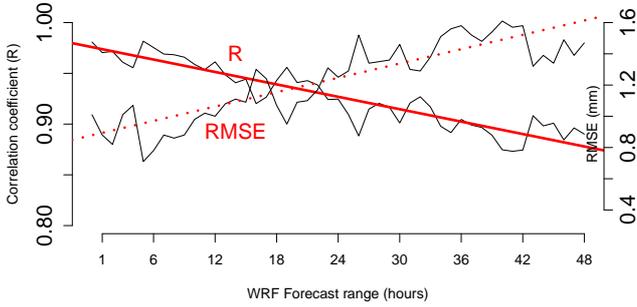} 
	\caption{Linear fits of the correlation between \PWVG~and \PWVW (solid line and left axis) and RMSE of the residuals \PWVG~$-$ \PWVW (dashed line and right axis) as a function of the WRF forecast range.}
	\label{fig:RMSE_R}
\end{figure}

\subsection{Methods}
\label{sec:methods}

There are basically two outputs of the WRF simulations in this study: the amount of PWV and the time stamps of the values. The first step in the validation is to compare these parameters with those measured by the local GNSS monitor. This comparison is carried out point by point for all the forecasting horizons, from $0$ to $48$ h. We then analyse the full capabilities of WRF as an operational tool running every $24$ h with a horizon of $48$ h by comparing \PWVG~with the series \PWVII~and \PWVIV~	in two ways: taking the whole series and subsampling the data as a function of the measured PWV.

For the validations we performed a linear regression analysis using \PWVG~as reference. The association between the two variables, \PWVW~and \PWVG, is obtained through the Pearson correlation coefficient ($R$). The final error associated with \PWVW~must include all the uncertainties in the validation:

\begin{equation}
	\label{eq:PWVW_error}
	\epsilon_{W}^2 = \rm{RMSE}_{res}^2 + \rm{RMSE}_{W,calib}^2 + \epsilon_{G}^2  ,
\end{equation}

\noindent where $\rm{RMSE}_{res}$ is the RMSE\footnote{The RMSE (Root-Mean-Square Error) is defined as the square root of the sum of the variance of the residuals and the squared bias.} of the residuals, which are defined as the difference between observations and forecasts (\PWVG~$-$ \PWVW), $\rm{RMSE}_{W,calib}$ is the RMSE of the calibration in eq.\ \ref{eq:calib_d03}, and $\epsilon_{G}$ is the error of \PWVG,

\begin{equation}
	\label{eq:PWVG_error}
	\epsilon_{G}^2 = \rm{RMSE}_{G,calib}^2 + \sigma_{G}^2  ,
\end{equation}

\noindent where $\rm{RMSE}_{G,calib}$ is the RMSE of the calibration in eq.\ \ref{eq:calib_gnss} and $\sigma_{G}$ is the median instrumental uncertainty ($0.92$ mm). The equations \ref{eq:calib_d03} and \ref{eq:PWVW_error} are valid both for \PWVII~and \PWVIV.

\begin{table}
 \centering
  \caption{Correlation ($R$) of \PWVG~(measured) and \PWVW~(forecasted), RMSE  of the residuals (\PWVG~$-$ \PWVW), and final error $\epsilon_{W}$ (Eq.  \ref{eq:PWVW_error}), from eq.\ \ref{eq:fit_R} and \ref{eq:fit_RMSE} (see also Fig.\ \ref{fig:RMSE_R}). $F_{range}$ is the forecast range.} 
 \label{symbols}
 \begin{tabular}{@{}lccr}
   \hline
   $F_{range}$ & R & RMSE &  $\epsilon_{W}$  \\
   (hours)     &  &  (mm) & (mm) \\
   \hline
   6  & 0.96 & 0.95 & 1.71 \\
   12 & 0.95 & 1.05 & 1.76 \\
   24 & 0.93 & 1.23 & 1.88 \\
   48 & 0.88 & 1.60 & 2.14 \\
   72 & 0.83 & 1.96 & 2.42 \\
   \hline
   \label{table:RMSE_R}
 \end{tabular}
\end{table}

\section{Results}
\label{sec:results}

Here we present and discuss the results of the validation described in Section \ref{sec:methods}. We first show the comparison of the WRF outputs, PWV, and time stamps. 

\subsection{WRF outputs performance: PWV}
\label{sec:pwv_performance}

Each daily execution of WRF generates 49 hourly forecasts with an increasing horizon from $0$ to $48$ hours (see Section \ref{sec:PWV_datasets}). In this section we have grouped all the WRF outputs into 49 time series as a function of such forecast horizons to compare with \PWVG. The results are plotted in Figure \ref{fig:RMSE_R}. The correlation, $R$, slowly decreases with the time horizon (from $\sim$0.97 to $\sim$0.88) with a linear trend. The linear least-squares fit gives the following equation:

\begin{equation}
\label{eq:fit_R}
R=-0.002\cdot F_{range} + 0.98,
\end{equation}

\noindent where $F_{range}$ is the forecast range in hours. The RMSE of the residuals also shows a slow linear increase with the forecast range (from $\sim$0.9 to $\sim$1.6 mm) with the following fit equation,

\begin{equation}
\label{eq:fit_RMSE}
RMSE=0.015\cdot F_{range} + 0.86.
\end{equation}

These results improved upon those obtained by \citet{gon13}, who reported RMSE of $\approx$1.6 mm and $\approx$2.0 mm, and correlation coefficients of $\approx$0.88 and $\approx$0.82 for 24- and 48-hour forecasts respectively, as well as for the mountains of the Canary Islands including data from LPAL. Different factors playing a role in these differences, such as the better resolution of the IBC from GFS in this study ($0.25\degr\times0.25\degr$), compared with $1\degr\times1\degr$ in \citet{gon13} and a more detailed WRF model configuration, among others.

These results allow us to assume that no significant degradation of the forecast is to be expected in longer time horizons, and we can extrapolate the correlation and RMSE for $72$ h, $R (72h) = 0.83$ and $RMSE (72h) = 1.96$ mm. Table \ref{table:RMSE_R} summarizes the main results interpolated from eqs \ref{eq:fit_R} and \ref{eq:fit_RMSE}.

\subsection{WRF outputs performance: time stamps}
\label{sec:timestamp_performance}

\begin{figure}
  	\includegraphics[width=84mm]{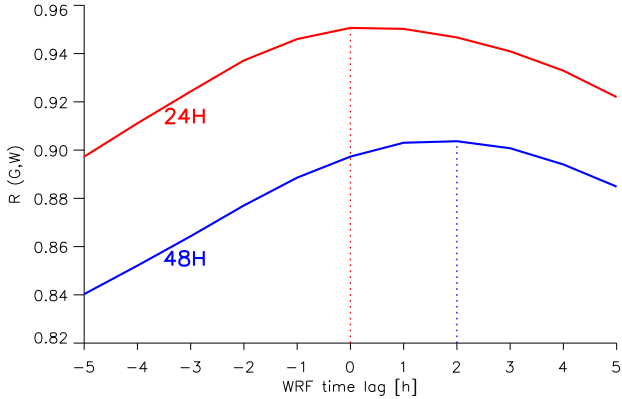}
	\caption{Correlation ($R$) of \PWVII~(red) and \PWVIV~(blue) with \PWVG~after applying time lags in steps of 1 hour. Positive (negative) time lags imply a forward (backward) shifting of the \PWVW~series.}
	\label{fig:WRF_lag}
\end{figure}

The time accuracy in the forecasts is also evaluated. A delay or advance when forecasting an abrupt change in the PWV content may increase the individual differences with the final values (i.e.\ residuals). We have analysed the WRF time accuracy in the operational series (\PWVII~and \PWVIV) calculating the loss in correlation after shifting the series in discrete steps of $1$ h (the time resolution in this study). The results are shown in Figure \ref{fig:WRF_lag}. Positive time lags imply a forward shift of the \PWVW~series. 

We found no time lags for the \PWVII~series, but we found one of about $2$ h for the \PWVIV~forecasting. This means that the $48$ h forecasts tend to be advanced in relation to the measured \PWVG. Therefore, \PWVIV~has to be corrected for this $2$ h time lag offset to achieve the maximum performance of the model. 

\subsection{Operational performance}
\label{sec:24_48h_timeseries}
\begin{figure*}
  	\includegraphics[width=180mm]{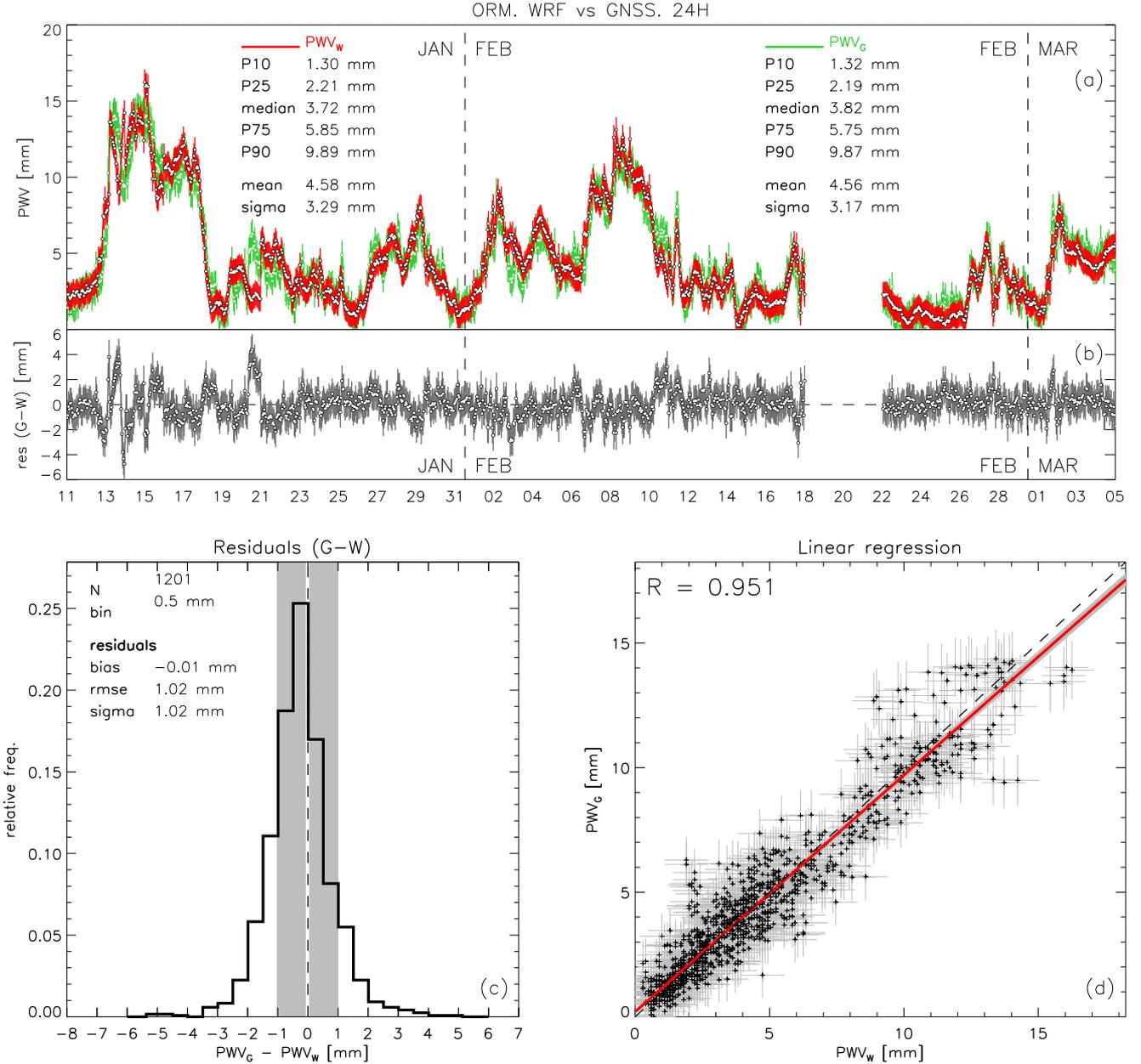}
	\caption{Statistical analysis of the full \PWVII~series (red) compared with the reference \PWVG~(green) (see Section \ref{sec:PWV_datasets}). The data time series are plotted in panel \emph{a}, the residual time series and distribution are in panels \emph{b} and  \emph{c}, and the regression analysis in panel \emph{d}.}
	\label{fig:PWV_24H}
\end{figure*}

\begin{figure*}
  	\includegraphics[width=180mm]{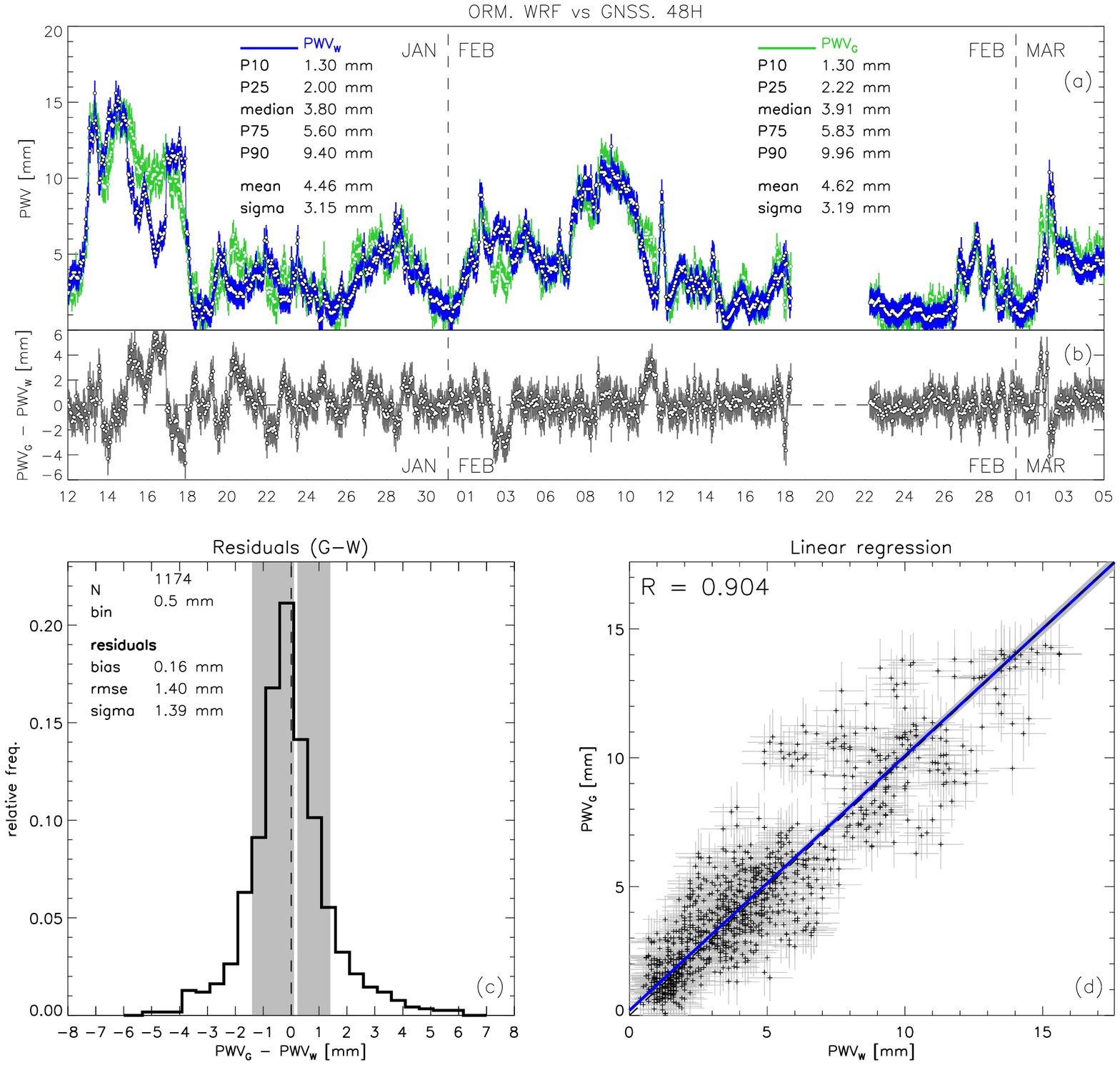}
	\caption{Statistical analysis of the full \PWVIV~series (blue) compared with the reference \PWVG~(green) (see Section \ref{sec:PWV_datasets}). The data time series are plotted in panel \emph{a}, the residual time series and distribution are in panels \emph{b} and  \emph{c}, and the regression analysis in panel \emph{d}.}
	\label{fig:PWV_48H}
\end{figure*}

The final performance of WRF as a valid operational tool for IR observations at the ORM is carried out through statistical analysis of the comparisons of \PWVII~and \PWVIV~with \PWVG. The results  are shown in the Figures \ref{fig:PWV_24H} and \ref{fig:PWV_48H}. The \PWVIV~series have been corrected for the $2$ h time lag described in Section \ref{sec:timestamp_performance}. Both figures (\ref{fig:PWV_24H} and \ref{fig:PWV_48H}, panels \emph{a}), show a wide range of PWV values and an excellent match of the measured and forecast series with time. The accuracy of the model is evaluated by the RMSE associated with the residuals, which are uniformly distributed about zero with a slight bias of $-0.01$ mm and RMSE $1.02$ mm for \PWVII, and $0.16$ mm and $1.40$ mm for the same parameters in \PWVIV~(see Figures \ref{fig:PWV_24H} and \ref{fig:PWV_48H}, panels \emph{b} and \emph{c}). The error in the forecast results from eq.\ \ref{eq:PWVW_error} with values of $1.75$ mm and $1.99$ mm for \PWVII~and \PWVIV~respectively. A good correlation is also reflected by the regression analysis (Figs \ref{fig:PWV_24H} and \ref{fig:PWV_48H}, panel \emph{d}) with Pearson correlation coefficients of $\emph{R}=0.951$ and $\emph{R}=0.904$. A summary of these results is given in Table \ref{table:stats}.

\subsection{WRF performance and PWV ranges}
\label{sec:performance_PWV_ranges}
\begin{figure}
  	\includegraphics[width=84mm]{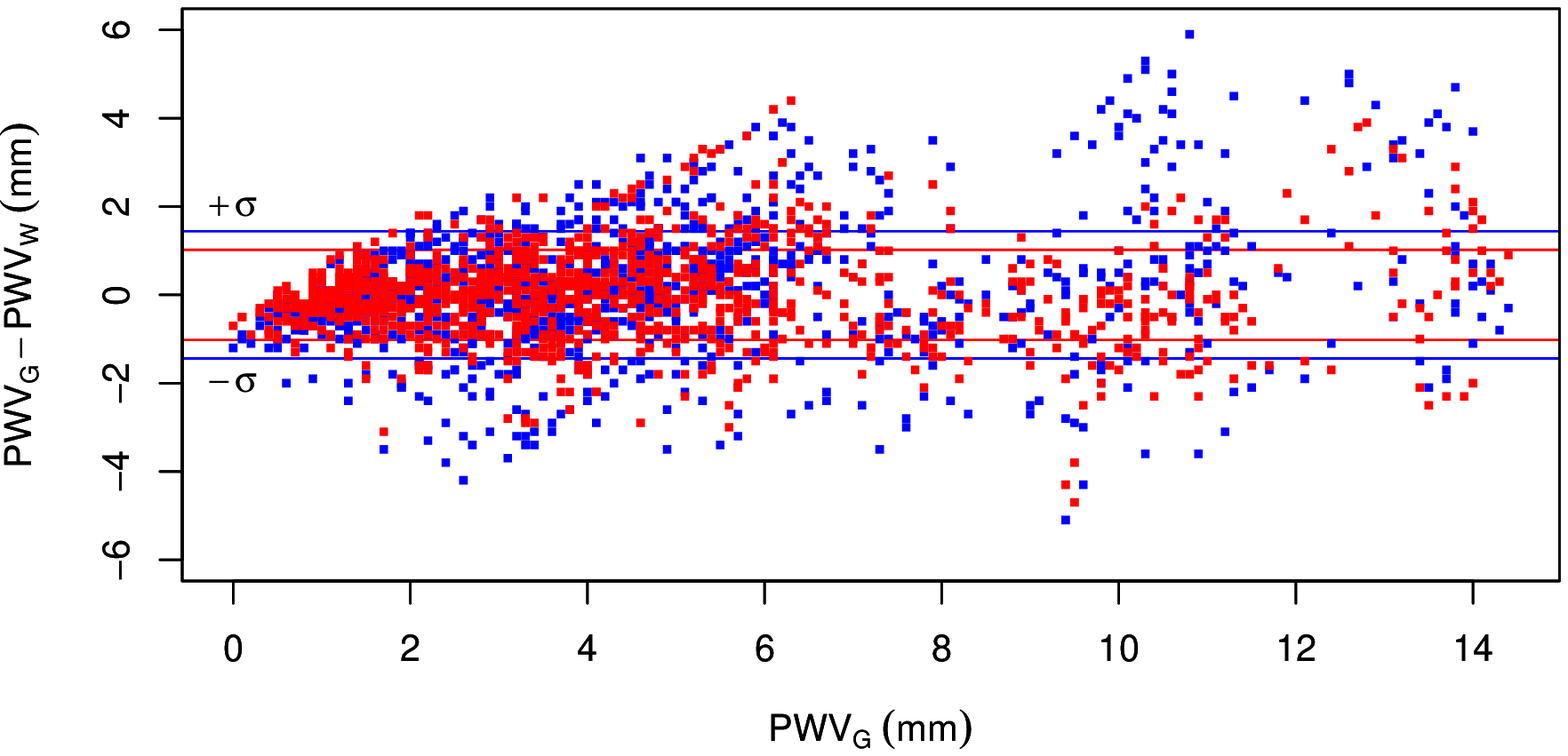}
	\caption{Dependence of the residuals for the $24$ h (\PWVG~$-$ \PWVII; red) and $48$ h (\PWVG~$-$ \PWVIV; blue) horizons. The horizontal lines are the standard deviation for both \PWVII~and \PWVIV, and follow the same colour scheme.}
	\label{fig:WRFres_PWV}
\end{figure}

\begin{figure}
  	\includegraphics[width=84mm]{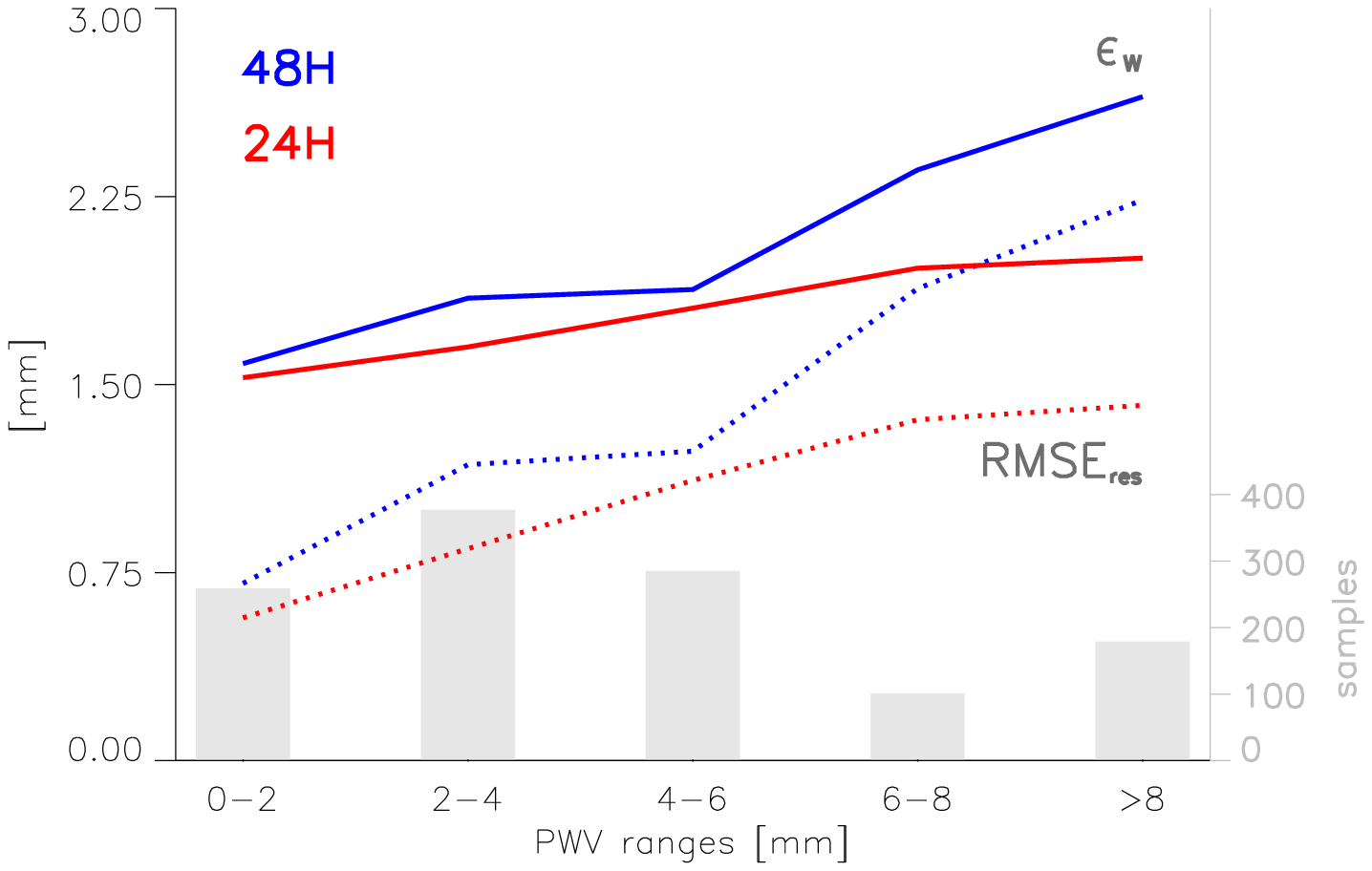}
	\caption{WRF performance for different PWV content (\PWVG). The solid lines are the errors ($\epsilon_W$) and the dotted lines are the RMSE of the residuals ($\rm{RMSE}_{res}$) for both the $24$ h (red) and $48$ h (blue) horizons. The light grey bars (right axis) are the number of samples for each bin.}
	\label{fig:PWV_ranges}
\end{figure}

The PWV was below $6$ mm for $\approx76$ \% of the period covered in the \PWVG~series. Different classifications have been proposed for the quality of IR observations as a function of PWV. For example, \cite{kid98} established a scale in which $0<$PWV$<3$ mm corresponds to good or excellent conditions, $3<$PWV$<6$ mm to fair or mediocre conditions, $6<$PWV$<10$ mm  to poor conditions, and PWV$>10$ mm to extremely poor conditions.

The WRF performance for different PWV values can be analysed through the behaviour of the residuals, as shown in Fig.\ \ref{fig:WRFres_PWV}, for both \PWVII~and \PWVIV~(see Section \ref{sec:PWV_datasets}). The residuals are more scattered as the PWV increases, with more dispersion in \PWVIV~than \PWVII, as expected. There is a slight \emph{wet} bias in the forecasts for the driest conditions (\PWVG~$\lesssim 1$ mm), reflected in negative residuals for this PWV range. Two factors may play a role in such an effect. On the one hand, the relative weight of small (below the horizontal resolution of the model) wet air pockets in the determination of the integrated PWV is larger for very dry conditions. On the other, the median error for the reference series (\PWVG) (see eq.\  \ref{eq:PWVG_error}) is $1.1$ mm, it being difficult to obtain conclusions below this value. A specific work with more accurate techniques would be required study of the WRF behaviour for PWV$<1$ mm.

To analyse the WRF performance for the different PWV values in more detail we have grouped the RMSE of the residuals and the final errors $\epsilon_W$ from eq.\ \ref{eq:PWVW_error} (for both \PWVII~and \PWVIV) in PWV ranges. The results are shown in Fig.\ \ref{fig:PWV_ranges} and in Table \ref{table:stats}. We constrained the analysis to the range 0--8 mm with a binning of $2$ mm. As in Fig.\ \ref{fig:WRFres_PWV},  Fig.\  \ref{fig:PWV_ranges} also reveals growth in the RMSE and errors with PWV, with better behaviour for \PWVII. 

\begin{table*}
 \centering
  \caption{Summary of the statistical validation of WRF for PWV forecasts. \PWVG~and \PWVW~are the time series of GNSS and WRF respectively (see Section \ref{sec:PWV_datasets}). The subscripts \emph{calib} and \emph{res} refer to the calibrations in eqs \ref{eq:calib_d03} and \ref{eq:calib_gnss}, and the statistics of the residuals (see Figures \ref{fig:PWV_24H} and \ref{fig:PWV_48H}). The time lag $\tau_{W}$ comes from Figure \ref{fig:WRF_lag}. The error and $\epsilon_{W}$ is obtained from eqs \ref{eq:PWVW_error} and \ref{eq:PWVG_error} with $\rm{RMSE}_{G,calib}=0.70$ mm,   $\sigma_{G}=0.92$ mm, and $\rm{RMSE}_{W,calib}=0.82$ mm. The values for different PWV ranges come from Section \ref{sec:performance_PWV_ranges}.}
 \label{table:stats}
 \begin{tabular}{@{}lc ccc ccr}
   \hline
   Forecast & PWV range & $\rm{bias}_{res}$  & $\sigma_{res}$ & $\rm{RMSE}_{res}$
                        & $\tau_{W}$         & $\epsilon_{W}$ & R\\ 

   horizon  & (mm)      & (mm)               & (mm)           & (mm)
                        & (h)                & (mm)           & \\
   \hline
   24 h \dots\dots & all   & -0.01 & 1.02 & 1.02 & 0 & 1.75 & 0.95  \\
                   & $0-2$ & -0.15 & 0.55 & 0.57 & 0 & 1.53 & 0.54  \\
                   & $2-4$ & -0.07 & 0.84 & 0.85 & 0 & 1.65 & 0.59  \\
                   & $4-6$ &  0.20 & 1.10 & 1.12 & 0 & 1.80 & 0.45  \\
                   & $6-8$ &  0.23 & 1.34 & 1.36 & 0 & 1.96 & 0.66  \\
                   &  $>8$ & -0.17 & 1.41 & 1.42 & 0 & 2.00 & 0.69  \\
\\
   48 h \dots\dots & all   &  0.16 & 1.39 & 1.40 & 2 & 1.99 & 0.90  \\
                   & $0-2$ & -0.24 & 0.66 & 0.71 & 2 & 1.58 & 0.32  \\
                   & $2-4$ & -0.12 & 1.17 & 1.18 & 2 & 1.84 & 0.46  \\
                   & $4-6$ &  0.54 & 1.11 & 1.23 & 2 & 1.88 & 0.34  \\
                   & $6-8$ &  0.17 & 1.87 & 1.88 & 2 & 2.36 & 0.51  \\
                   &  $>8$ &  0.70 & 2.12 & 2.24 & 2 & 2.65 & 0.55  \\
   \hline
   \label{table:stats}
 \end{tabular}
\end{table*}

\section{WRF performance for abrupt PWV gradients}
\label{sec:PWV_gradients}

Episodes of steep PWV gradients occurred in the period covered in this study (see Figs \ref{fig:PWV_24H} and \ref{fig:PWV_48H}), although such episodes may be considered unusual. In fact, the median PWV for the \PWVG~series is $3.82$ mm, slightly above the value of $2.9$ mm reported for winter at the LPAL station by \cite{gar10}. This climatological scenario allowed us to test the WRF forecasts for a wide range of meteorological conditions at ORM, including sharp and abrupt changes.

Both series of WRF outputs, \PWVII~and \PWVIV, were able to forecast all the steep gradient events. The only exception is an episode between January 20 and 21, when the \PWVW~was uncorrelated with \PWVG~for some hours.

\subsection{Case study: January, 20-21, 2016}
\label{sec:jan20}

\begin{figure}
  	\includegraphics[width=84mm]{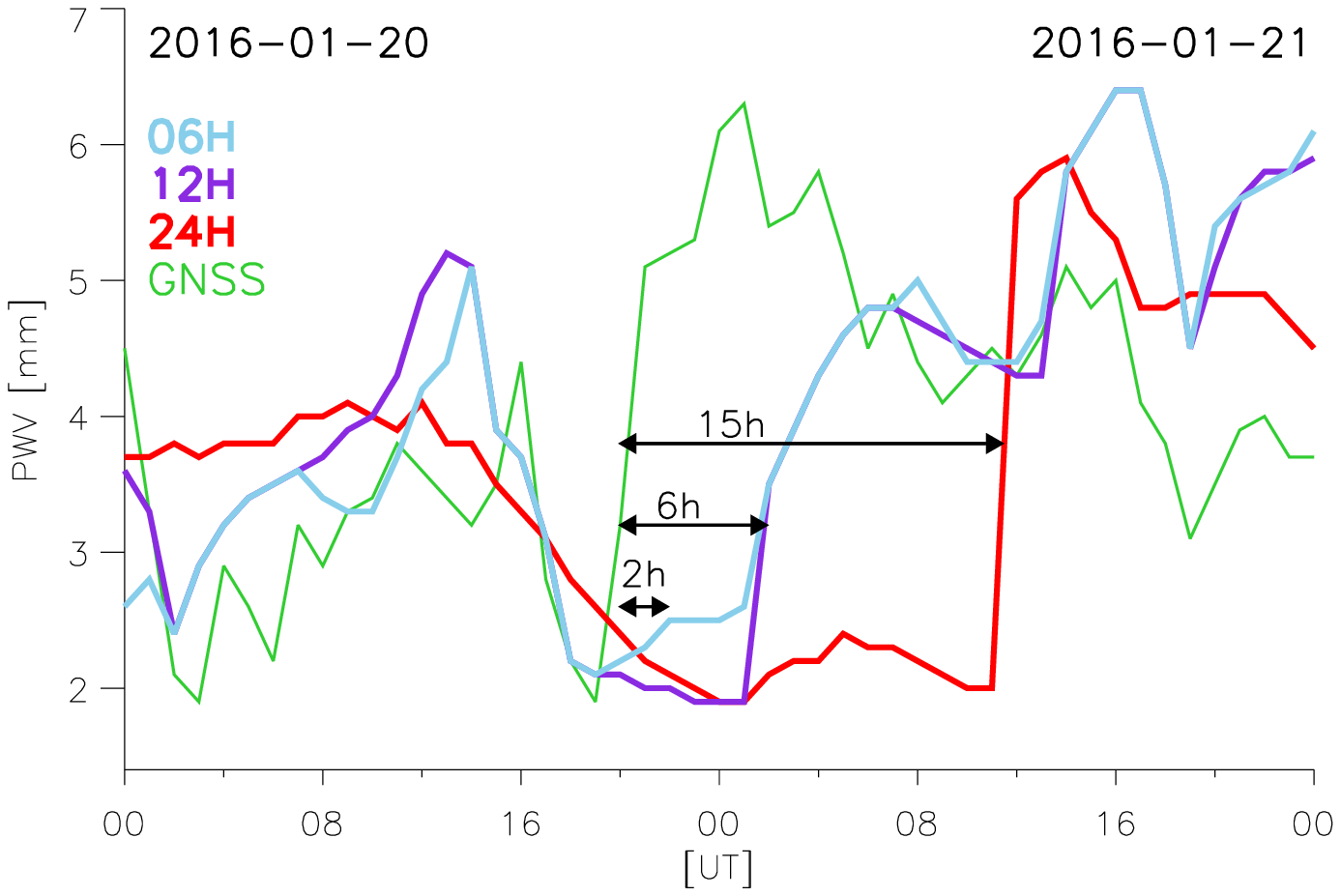}
	\caption{\PWVG, \PWVII, PWV$_{\rm{W12}}$, and PWV$_{\rm{W06}}$ time series for the event between January 20 and 21 (see colour codes in the legend). The horizontal arrows and labels indicate the delay in the WRF response to the steep increase in PWV, as measured by the GNSS monitor. See Sec. \ref{sec:jan20} for details.}
	\label{fig:jan20}
\end{figure}

A singular situation took place in $\approx24$ h period between January, 20 and 21.  A very pronounced delay in the \PWVW~forecasted signal was observed in the time series over a fluctuation of $4.4$ mm in \PWVG~(see panel \emph{a} of Figure \ref{fig:PWV_24H}). To assess the origin of such a delay, we ran WRF specifically for this episode with higher frequencies of $12$ and $6$ h (PWV$_{\rm{W12}}$ and PWV$_{\rm{W06}}$, hereafter); that is, with more frequent updates of the IBC. All the series for this period (\PWVG, \PWVII, PWV$_{\rm{W12}}$, and PWV$_{\rm{W06}}$) have been plotted in Figure \ref{fig:jan20}. There is a clear improvement when increasing the frequency of the WRF runs, with a reduction in the initial delay of $\sim15$ h (\PWVII) to $\sim6$ h for PWV$_{\rm{W12}}$. In the following step, PWV$_{\rm{W06}}$, WRF is able to forecast the increase of PWV $\sim4$ h in advance, but with an inaccurate value. Although this event is isolated, these results seem to show that some PWV features may be extremely 
local, and that WRF therefore becomes limited by the spatial resolution. A more detailed study of these phenomena is beyond the scope of this paper.

\section{WRF as an operational forecasting tool for PWV}
\label{sec:operational_tool}

In the context of operational forecasting of PWV, the WRF model could  currently be run up to four times a day using the available operational GFS model outputs at 00, 06, 12, and 18 UTC. The total execution time for a single simulation is the sum of the pre-processing, WRF execution, and WRF output post-processing and generation of final products. In a typical Linux machine with $\approx$12 cores, it lasts between 2 and 4 hours. The desired horizontal and vertical resolution, the extent of the domains, and the forecast range severely impact on the computing time, so these parameters must be selected properly in line with the operational requirements of user telescopes. The WRF architecture supports parallelization, so the program can be run in a Linux cluster with multiple CPUs, thereby significantly reducing the execution time.

\section{Conclusions}
\label{sec:conclusions}

The WRF model has proven to be very good at predicting PWV above the ORM up to a forecast range of 48 hours. Our main findings are summarized as follows:

\begin{itemize}
    \item[$-$] Excellent agreement between model forecasts and observations was found with $R = 0.951$ and $R = 0.904$ for \PWVII~and \PWVIV, respectively.
    \item[$-$] The total PWV forecast errors are $1.75$ mm for \PWVII~and $1.99$ mm for \PWVIV.
    \item[$-$] We found linear trends in both the correlation and RMSE of the residuals as a function of forecast range. The RMSE slowly increases with the forecast range (ranging from $\sim$0.9 to $\sim$1.6 mm ), whereas the correlation between observations and the forecasts decreases (from $\sim$0.97 to $\sim$0.88).
    \item[$-$] Assuming a linear trend, the extrapolated forecast error up to $72$ h is $2.4$ mm. 
    \item[$-$] The PWV amount impacts on the forecast performance with slow growth in the RMSE as the PWV increases. \PWVII~behaves better than \PWVIV~for all PWV ranges.
    \item[$-$] The time accuracy in the forecasts impacts on  model performance. No time lags were found for the \PWVII~series, but a time lag of $2$ h was present for \PWVIV.
    \item[$-$] WRF was able to trace all sudden changes in PWV on short timescales except for one case, for which a higher temporal resolution would be necessary.
    \item[$-$] Besides its operational use as a forecasting tool, the accuracy of the WRF forecasting tool for PWV allows it to be used as a backup of the real-time GNSS PWV monitor in case of failure. 
    \item[$-$] In summary, the WRF performance is excellent and accurate, allowing it to be implemented as an operational tool at the ORM with horizons of $24$ and $48$ h.
\end{itemize}

\section*{Acknowledgments}
\label{sec:ack}
This study has been funded by the Instituto de Astrof\'isica de Canarias (IAC). The GNSS PWV monitor at the ORM was developed by the IAC with the subcontractor Soluciones Avanzadas Canarias. The WRF is maintained and supported as a community model and can be freely downloaded from the WRF user website\footnote{http://www.mmm.ucar.edu/wrf/users/}. WRF model version 3.1 was used in the present study. We would like to acknowledge the scientific community supporting WRF. The GFS model is provided and run by the National Centers for Environmental Prediction (NCEP). We would also acknowledge them for access. Special thanks are given to Terry Mahoney (IAC) for the English language correction.

\bibliography{ipwv_paper}
\bibliographystyle{mnras}

\end{document}